\documentclass{icrc2009}

\usepackage{graphicx}   % for including figures
\usepackage[caption=false]{caption}    % for captions
\usepackage[font=footnotesize]{subfig} % subfig.sty for a double column floating figure using two subfigures
\usepackage{fixltx2e}
\usepackage{url}

\newcommand{\shorttitle}[1]%
{\markboth{Proceedings of the 31\MakeLowercase{$^{st}$} ICRC, {\L}\'{o}d\'{z} 2009}{#1} }
\newcommand{\etal}{\MakeLowercase{\textit{et al. }}} % "et al."

\newcommand{\japj}{Astrophys. J.}

\newcommand{\jasr}{Adv. Space Res.}
\newcommand{\jprep}{Phys. Rep.}

\newcommand{\jarXiv}[1]{arXiv:#1}

\newcommand{\jnima}{Nucl. Instrum. Methods Phys. Res. A}

\newcommand{\jtbp}{(to be published)}
\newcommand{\jloc}[3]{{\bf #1}, #2 (#3)}
\newcommand{\jref}[4]{#1, \jloc{#2}{#3}{#4}}

\begin{document}

\title{First results on Cosmic Ray electron spectrum below 20 GeV from the Fermi LAT.}

\author{\IEEEauthorblockN{M.Pesce-Rollins\IEEEauthorrefmark{1} on behalf of the Fermi LAT Collaboration}\\
\IEEEauthorblockA{\IEEEauthorrefmark{1}Istituto Nazionale di Fisica Nucleare, Sezione di Pisa,I-56127 Pisa, Italy}}

\shorttitle{M.Pesce-Rollins \etal CR electrons below 20 GeV}
\maketitle

\begin{abstract}
Designed to be a successor of the previous flown space based 
gamma ray detectors, the Fermi Large Area Telescope (LAT) is 
also an electron detector. Taking advantage of its capability 
to separate electromagnetic and hadronic signals it is possible to accurately 
measure the Cosmic Ray electron spectrum. The spectra of primary cosmic ray 
electrons below 20 GeV is influenced by many local effects 
such as solar modulation and the geomagnetic cutoff. For energies below a 
few GeV it is possible to observe the albedo population of electrons 
which are controlled by the local magnetic field. In this paper
we present the LAT electron analysis in particular event selection and 
validation as well as the first results on the measurement of the 
electron spectrum below 20 GeV.
\end{abstract}

\begin{IEEEkeywords}
Cosmic Ray electrons, geomagnetic cutoff, Fermi Large Area Telescope
\end{IEEEkeywords}
 
\section{Introduction}

High-energy Cosmic Ray (CR) electrons are believed to orginate from primary
acceleration sites such as supernova remnants and pulsars. Electron-position
pairs are thought to be produced from the collisions of CR hadrons and gamma
rays with the interstellar gas. An accurate measurement of the
CR electron\footnote{In this paper we will refer to electrons as the sum of $e^+ + e^-$ unless specified otherwise} spectrum can help unveil the origin of these 
particles as well as their propagation through the interstellar medium. By 
measuring the CR electron spectrum below 20 GeV we have the unique
possibility to study not only the primary CR spectra but also the behavior
of atmospheric secondaries in the upper atmosphere. Prior to 2008, the electron
spectrum was measured by balloon-borne experiments~\cite{HEAT} and by a single space mission
(AMS-1)~\cite{AMS}. These pioneering experiments have made important contributions to the
understanding of the electron spectra. Nonetheless in order to have a full
comprehension of the distribution of the secondary particles it is of 
fundamental importance to sample this population over the widest energy 
range and geomagnetic position as possible. Futhermore, the measurement of 
the primary electron spectra down to the geomagnetic cutoff is necessary to 
constrain many CR propagation models. In this paper we present the CR 
electron spectrum measured by the Fermi Large Area Telescope (LAT) in the 
energy range below 20 GeV.
\section{The Fermi LAT}
The Large Area Telescope (LAT) is the main instrument on-board the recently 
launched Fermi Gamma-Ray Space Telescope mission. The Fermi satellite 
is in a circular orbit at 565 km altitude with a 25.6$^{\circ}$ inclination.
The LAT is a modular array of 4$\times$4 \emph{tower} modules, each one including a tracker/converter (TKR),surrounded by an Anti-Coincidence Detector (ACD), a calorimeter (CAL) and an electronics module. The TKR takes advantage of the Silicon Strip Detector (SSD) technology, allowing a high precision tracking with a very small dead time. Each of the 16 CAL modules is composed of 96 CsI crystals, for a total depth of 8.6 radiation lengths, arranged in a hodoscopic configuration. This allows for a full 3D reconstruction of the induced shower and helps in the discrimination between electromagnetic and hadronic showers as well as leakage correction at high energy. The ACD is composed of 89 independent plastic scintillator tiles. The information from each of these tiles is available on-board for the high level trigger as to limit the effect of self veto at high energies~\cite{LATpaper}. Given the fact that the Fermi LAT is a pair conversion telescope it is by its nature also an electron positron detector.

The orbital parameters of the Fermi mission define the geomagnetic latitude 
range over which we can sample the primary CR electron spectrum and also the 
lower limit on the geomagnetic cutoff value. Due to the fact that most of the existing data from previous experiments was taken in different orbits it is convinient to organize the data in L shell bins to facilitate comparison. For a dipole field, 
\begin{equation}
R = R_0\cos^2\theta
\end{equation}
defines lines of constant B, also known as "L shells". ${\rm R_0}$ is the radial distance to the field line where it crosses the geomagnetic equator and R is the radial distance to the point where the field is B at latitude $\theta$. L is then defined as ${\rm L} = {\rm R_0}/{\rm R_E}$ where ${\rm R_E}$ is the Earths's radius (6371 km). Due to the fact that the Earth's dipole is offset and tilted with respect to the center of the Earth, L values must be calculated based on a detailed field model and are a function of geomagnetic latitude~\cite{GeoMagneticVars}. 
With Fermi's orbit we are able to measure the spectrum in the range 1.0 $<$ L $<$ 1.73. No data is taken while passing through the South Atlantic Anomaly (SAA). The fraction of time that Fermi spends in the SAA is 14.6\%. 
\section{Analysis}
\begin{figure*}[!t]
\centerline{\subfloat[Geometry factor]{\includegraphics[angle=90,width=0.5\textwidth]{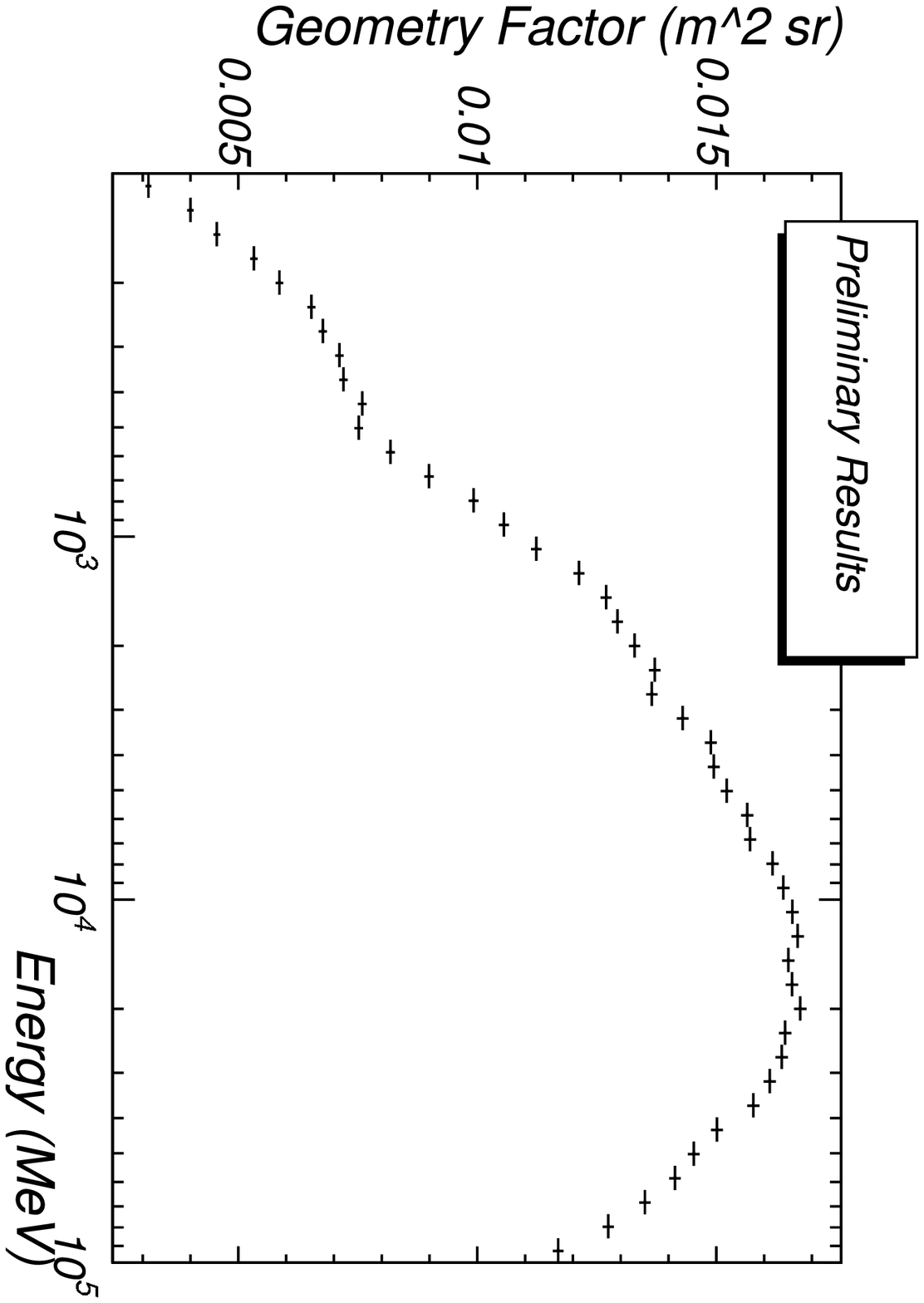}\label{GF}}\hfil
\subfloat[Orbital Averaged Contamination]{\includegraphics[angle=90,width=0.5\textwidth]{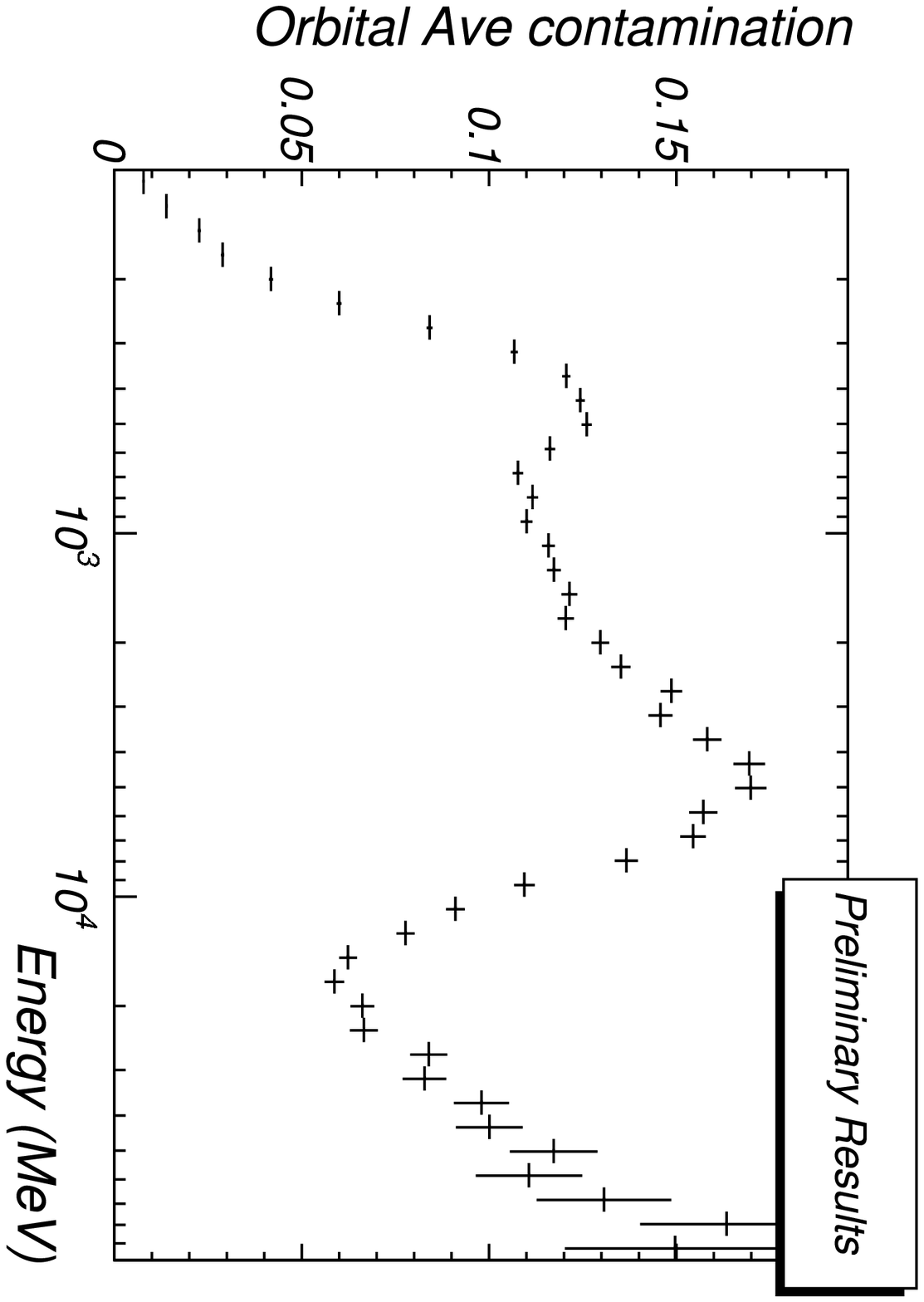}\label{Contamination}}}
\caption{The Geometry factor (GF) for the electrons (a). The 1:250 prescale factor from the DGN filter is already taken into account in the calculation of the GF. The orbital averaged residual hadron contamination after the event selection cuts (b).}
\label{GF_Contam}
\end{figure*}
As described in~\cite{FermiElectrons}, the main on-board filter is configured to accept all events that deposit at least 20 GeV in the calorimeter. There is however another on-board filter, the Diagnostic filter (DGN), which is configured to pass a prescaled (1:250) unbiased sample of all the events that trigger the LAT. This filter was conceived, as the name implies, for diagnostic purposes such as the monitoring of the background rates and the characterization of the efficiency of the other on-board filters. The average downlink rate from the DGN filter is $\sim$ 20 Hz~\cite{LATpaper}. Thanks to the DGN filter it is possible to measure the electron spectrum with the Fermi LAT below 20 GeV\footnote{The sample of events that are downlinked via the DGN is not limited to energies below 20 GeV, but it serves as a complement to the main on-board filter.}.
Even though the DGN filter has a 1:250 prescale applied on-board, the fact that below a few GeV the electrons have very large fluxes (from hundreds of particles m$^{-2}$s$^{-1}$ to several tens of particles m$^{-2}$s$^{-1}$)~\cite{LATpaper} compensates for the prescale and allows us to gather large statistics (more than 5$\times$ 10$^6$ electron candidates in the first six months) through this filter.
 
Two very important figures of merit that come out of our dedicated event analysis are the geometry factor and residual hadron contamination. The geometry factor (GF) is the instrument acceptance, i.e. the product of the instrument field of view and effective area after applying selection 
criteria; the hadron contamination is the fraction of residual hadrons over 
all selected events. The GF and orbital averaged hadron contamination after the selection cuts have been applied can be seen in figure \ref{GF_Contam}. The residual hadron contamination is maintained well below 20\% throughout the entire energy range in question.
The fraction of electrons with respect to all the particles triggering the LAT (prior to the selection cuts) 
has very large variations with energy, and therefore it is imperative to 
optimize the selection cuts as a function of energy. This process needs to 
be carefully balanced between two requirements, namely to obtain high 
electron efficiency while not introducing any artificial feature in the 
resulting spectrum. Another important requirement of the selection cuts is the
hadron rejection power needed to be able to obtain a low residual contamination
in the final measured spectrum. Based on the data taken from past experiments on CR, splash and reentrant particles~\cite{LATpaper} we are able to estimate the energy dependent hadron rejection power needed for our study, namely from 1:10$^2$ for energies below $\sim$ 1 GeV to 1:10$^3$ for the high energy range.
The event selection developed for the electrons, just like for the photons~\cite{LATpaper}, 
relies on the LAT capability to discriminate EM and hadronic showers based on 
their longitudinal and lateral development, as measured by both the TKR and the
CAL detectors. The overall approach to the event selection can be separated into several steps each of which is designed to address different energy dependent requirements. 
First we apply a set of \emph{quality} cuts that serve to select all those events that are classified as well reconstructed as well as having failed the ACD vetoes developed for the background rejection for the standard photon analysis~\cite{LATpaper}.
The ACD is used in conjunction with the found tracks in order to determine on-board whether an event is a charged particle or a neutral. This approach is used to remove the vast majority of the background that enters within the Field Of View (FOV) in the photon analysis and is used in the electron analysis to remove the neutral events by taking all those events that fail this veto.
The \emph{quality} cuts  have an overall hadron rejection power of $\sim$1:10. 

The next steps consist in a detailed analysis of the event topologies in the LAT's three sub detectors, for example the different distributions of energy in the ACD, between EM and hadron initiated showers in the CAL or the distribution of TKR clusters around the main reconstructed track. These combined steps have the resulting rejection power of about 1:10$^2$. 
\begin{figure*}[!t]
\subfloat[Electron spectra in three McIlwain L bins]{\includegraphics[width=0.5\textwidth]{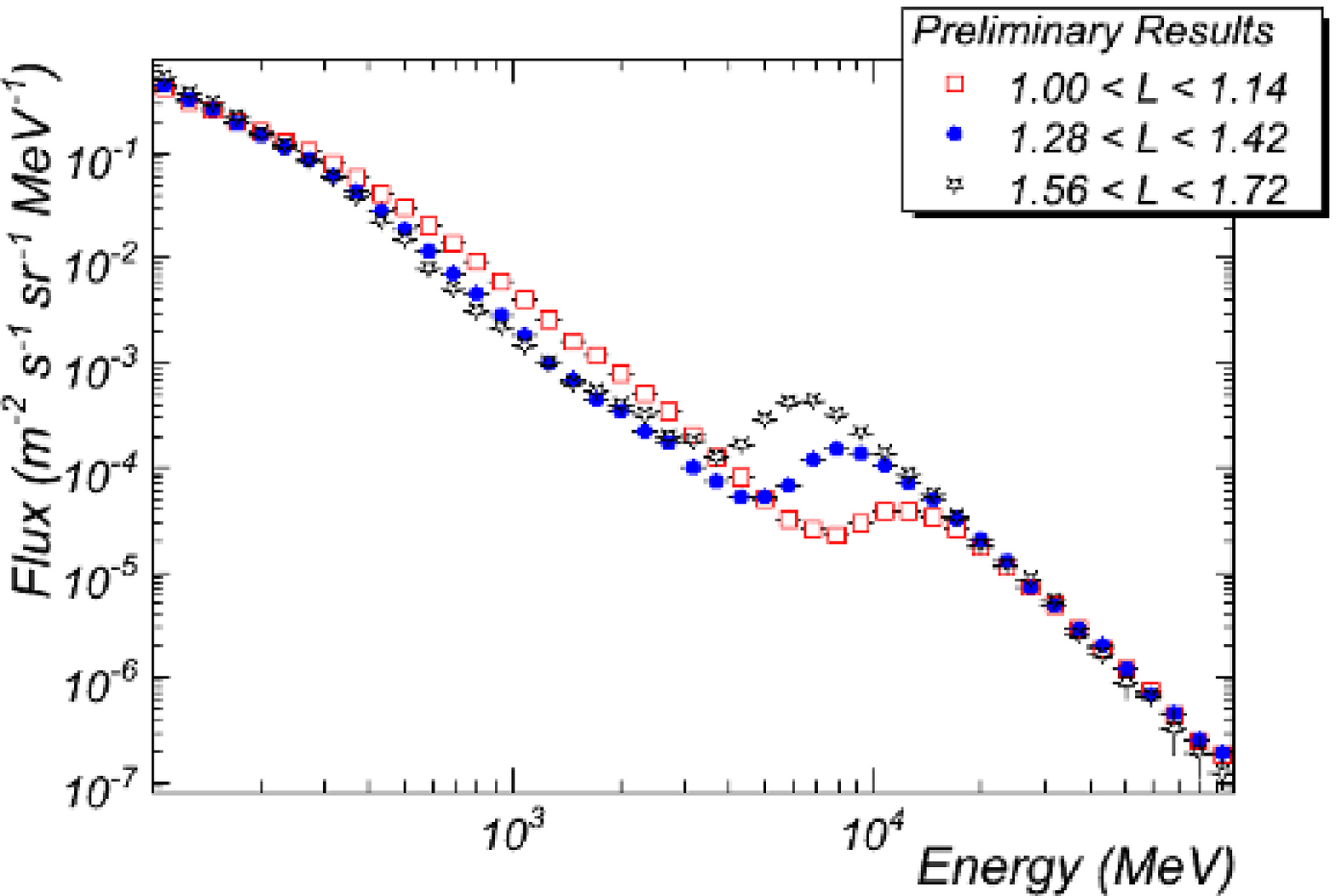}\label{Flux}}\hfil
\subfloat[East-West asymmetry in electron spectra]{\includegraphics[width=0.5\textwidth]{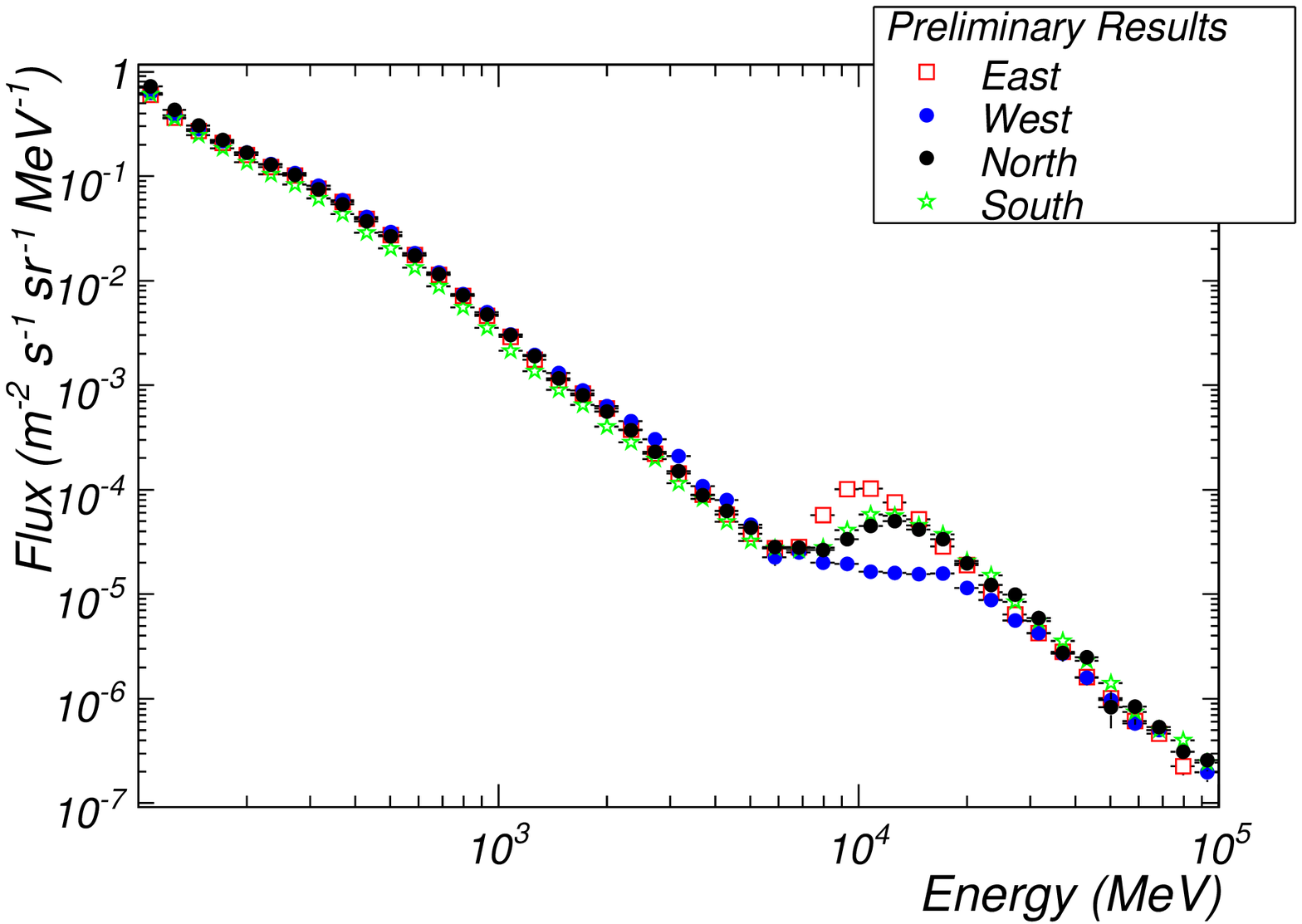}\label{EastWest}}
\caption{CR electron spectra in three different McIlwain L bins. The geomagnetic cutoff clearly shifts towards lower energy values with increasing McIlwain L position. Only three L bins are shown here for graphical clarity, the study was performed on a varying number of bins ranging from four to six (a). Clear signs of east-west symmetries in the electron spectra. Red squares correspond to the intensities originating from the east and blue circles from the west (b). These fluxes are preliminary. }
\label{Flux}
\end{figure*}
The remaining necessary boost in the rejection power is obtained via several
probability variables which result from training classification trees (CT)
to distinguish between EM and hadron events in the LAT subdetectors. It is important to note here that the CT probability variables used in this analysis come from two different sets of training, the first comes from the standard Fermi 
photon analysis and the second was performed especially for the electron analysis. 
This is done using large sets of Monte-Carlo (MC) events generated by the 
accurate LAT simulation package based on the \texttt{GEANT4} toolkit~\cite{Geant4}. The 
classifiers allow selection of the electrons through a multitude of parallel 
paths, each with different selections, that map the many different topologies 
of the signal events into a single, continuous probability variable that is used to simultaneously handle all valid selections. 
The residual hadronic background was taken as 1- \emph{purity}. Where the \emph{purity} was estimated from the fraction of electrons that survive the selection cuts over the total number of initial events in the MC simulation. All the candidate electrons passing the selection cuts are multiplied by this fraction in order to remove the expected background contamination. The selection cuts have been optimized on an orbital averaged MC population of events. However, it is clear that the background varies with energy as well as with orbital position. The optimization of the event selection and background subtraction as a function of L still needs to be performed. It is also important to note that the selection cuts were optimized using pre-Fermi background models and therefore the estimated hadron contamination may vary due to the uncertainties tied to the particle fluxes. One of the goals of this analysis is also aimed at updating the current knowledge of the electron fluxes in the cosmic ray models.

\begin{figure*}
\centering
\includegraphics[width=15.5cm]{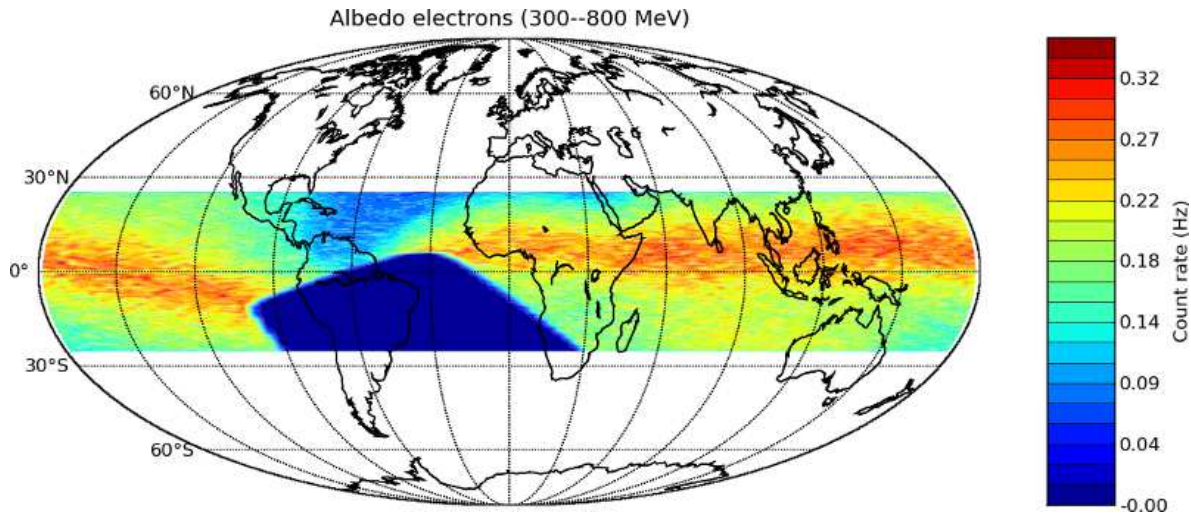}\label{sub_fig2}
\caption{Distribution of the secondary electrons (300 MeV $<$ E $<$ 800 MeV) mapped over the globe. Please note that these are count rates and not fluxes, the distribution of the fluxes is still work in progress and these rates are preliminary. The highest concentration of these trapped particles is seen in the magnetic equitorial region.}
\label{Globe}
\end{figure*}
\section{Results}
The electron spectrum for E less than $\sim$ 20 GeV has two main 
sources, the primary galactic cosmic rays and the secondary cosmic rays. 
Where for secondary we intend all the electrons that are produced from 
interactions of incident cosmic rays in the atmosphere. 
The high statistics (more than 5$\times$ 10$^6$ electron candidates) gathered just in the first six months of data taking enable us to study the variation of the spectra both above and below the geomagnetic cutoff as well as at varying L values. 
As can be seen from figure \ref{Flux} the geomagnetic cutoff position in energy clearly decreases with increasing values of L as expected. The spectrum above cutoff is seen to drop off according to a power law with spectral index of -3.04~\cite{FermiElectrons}. Only three L bins are shown in figure \ref{Flux} for graphical clarity, but the study was performed on a varying number of bins ranging from four to six. Also in figure \ref{Flux} it is possible to see east-west asymmetries in the flux in the geomagnetic cutoff region. These asymmetries are expected due to the fact that the magnetic rigidity is not only a function of orbital position but also of the particle charge. Due to this dependancy, when the primary electrons reach the earth and interact with the geomagnetic field they are deflected towards the west and therefore manifesting a larger intensity from the east (vice versa for the positrons). Since there are more primary electrons than primary positrons it is clear why we see such a large separation between the intensities originating from the east as opposed to those coming from the west. These fluxes are preliminary.

The reentrant albedo (secondary) electron spectrum as measured by Fermi also varies with L. Figure \ref{Globe} illustrates the distribution of the count rate for the electrons with reconstructed energy between $\sim$ 300 MeV and $\sim$ 800 MeV. From this figure it can be seen that there is a maximum concentration of trapped electrons in the geomagnetic equatorial region. Such properties were also reported by AMS precursor mission flown on the STS-91 in June 1998~\cite{AMS}. In  fact in~\cite{AMS} they associate such properties to the so-called \emph{long lived} leptons, leptons with flight times $\geq$ 0.2 sec. Through the aid of particle tracing methods it was possible to conclude that these electrons are produced and absorbed in the equitorial region.  

\section{Acknowledgements}
The Fermi LAT Collaboration acknowledges support from a number of agencies and institutes for both development and the operation of the LAT as well as scientific data analysis. 
These include NASA and DOE in the United States,\newpage  CEA/Irfu and IN2P3/CNRS in France, ASI and INFN in Italy, MEXT, KEK, and JAXA in Japan, and the K.~A.~Wallenberg Foundation, 
the Swedish Research Council and the National Space Board in Sweden. Additional support from INAF in Italy for science analysis during the operations phase is also gratefully acknowledged.


\begin{thebibliography}{99}
\bibitem{HEAT}
M. A. DuVernois \etal, \jref{\japj}{559}{296}{2001}.

\bibitem{AMS}
M.~Aguillar \etal, \jref{\jprep}{366}{331}{2002}.

\bibitem{LATpaper} W.~Atwood \etal, \japj\ \jtbp;
\jarXiv{0902.1089}.

\bibitem{GeoMagneticVars}
D.F. Smart, M.A. Shea \jref{\jasr}{36}{2012-2020}{2005}

\bibitem{FermiElectrons}
A.~A.~Abdo \etal, \emph{Accepted for publication in PRL on April 21, 2009}.

\bibitem{Geant4}
S.~Agostinelli \etal, \jref{\jnima}{506}{250}{2003}.


\end{thebibliography}
\end{document}